\documentclass[aps,epsfig,amstext, showpacs, floatfix]{revtex4}
\usepackage{graphicx}
\usepackage{amssymb}

\def\Journal#1#2#3#4{{#1} {\bf #2}, #3 (#4) }

\def\CJP{Can.J.Phys}

\def\EPJC{{Eur.Phys.J.}C}

\def\JHEP{J.High.En.Phys.}
\def\JPG{{J. Phys.} G}
\def\JPE{{J. Phys.} E}

\def\NAT{Nature}

\def\NCA{{Nuovo Cimento} A}
 
\def\NIMA{{Nucl. Instrum. Methods} A} 
\def\NPA{{Nucl. Phys.} A}

\def\NPB{{Nucl. Phys.} B}
\def\NPPS{{Nucl. Phys. Proc. Suppl.} B}

\def\PLB{{Phys. Lett.} B}

\def\PL{{Phys. Lett.}}

\def\PR{Phys. Rev.}
\def\PRL{Phys. Rev. Lett.}

\def\PRD{{Phys. Rev.} D}

\def\INTA{{Int. J. Mod. Phys.} A}
\def\EPJ{Eur.Phys.J.}

\def\g{$(g-2)$}

\newcommand{\be}{\begin{equation}}
\newcommand{\ee}{\end{equation} \noindent}

\begin{document}
\baselineskip=7mm
\bibliographystyle{unsrt}
\title{A new ring structure for muon \g\ measurements}
\author{F.J.M.~Farley\footnote{Address for correspondence: 8, Chemin de Saint Pierre, 06620 Le Bar sur Loup, France}}

\affiliation{Department of Physics, Yale University, New Haven, Connecticut 06520}

\date{\today}

\begin{abstract}
In this storage ring of discrete magnets with uniform field and edge focusing, the field averaged over the orbit is independent of orbit radius (particle momentum).  This ring is suitable for measuring the anomalous magnetic moment of the muon.  The field, averaged over the orbit,  can be calibrated by injecting horizontally polarized protons of the same momentum and measuring the proton $(g-2)$ precession.  The experiment then measures the ratio of muon to proton anomalous moments.  A measurement of muon (g-2) at the level of $0.03~ppm$ appears feasible.

\vspace{4mm}
Keywords: muon, magnetic moment, g-2, storage ring
\end{abstract}

\pacs{13.40.Em, 12.20.Fv, 14.60.Ef, 29.20.Dh}

\maketitle

\pagebreak

The anomalous magnetic moment of the muon $a \equiv (g-2)/2$ is sensitive to physics beyond the standard model \cite{recth}.  Recent experimental results \cite{benn02} with an error of $0.7~ppm$ (parts per million) are above the latest theoretical evaluations \cite{dave02,jege02,hag03,ghoz03} and many papers have been published commenting on this apparent discrepancy.  The theory depends on experimental data for hadron production in $e^+e^-$ collisions obtained with  electron-positron storage rings \cite{akhm00,bai00}, which are steadily being improved.   If the discrepancy persists, and if the remaining uncertainties in the theory can be reduced, it could be interesting to make a more precise measurement.

In the experiment, polarized muons, with charge $e$ and mass $m$, are traditionally trapped to make many turns in a magnetic storage ring with field $B$.  The spin then turns faster than the momentum vector with difference frequency 
\be
f_a^{\mu} \ =\ a\, (eB/2\pi mc) \label{eq:g2}
\ee
while the precession frequency of muons at rest in the same field is
\be
f_s^{\mu}\ =\, (1+a)(eB/2\pi mc)\ = \lambda f_s^p   \label{eq:fs}
\ee
The field averaged over the muon orbits is usually measured by the corresponding proton magnetic resonance (NMR) frequency $f_s^p$ and the precession frequency for muons at rest in the same field $f_s^{\mu}$ is then derived from the ratio of magnetic moments
\be
 \lambda = \mu ^{\mu}/\mu ^p = f_s^{\mu}/f_s^p \label{eq:lam}
 \ee
 which is known \cite{liu99} to $31~ppb$ (parts per billion).
 
Combining (\ref{eq:g2}), (\ref{eq:fs}) and (\ref{eq:lam}) the anomaly is calculated from
\be
a = \frac{R}{\lambda - R} \label{eq:g2calc}
\ee
where $R=f_a^{\mu}/f_s^p$ is the ratio measured in the experiment.

Traditionally the magnetic field is mapped by proton NMR with static probes which are gradually moved around the ring.  This requires the ring magnet to be continuous in azimuth without straight sections; large gradients at the end of discrete magnet segments would kill the signal.  Furthermore it is difficult to locate the muon orbits precisely in radius, so it is desirable for the mean field to be independent of orbit radius.  The first muon storage ring at CERN \cite{bail72} was weak-focusing, with the field changing by $50~ppm$ for a $1~mm$ increment in radius.  Fractional $ppm$ measurements cannot be made in such a ring.  Subsequent experiments \cite{bail79, benn02} therefore used a uniform magnetic field with vertical focusing supplied by electric quadrupoles.  It was shown \cite{bail77} that if the muons are stored at $3.096~GeV$, called the ``magic energy'', the radial electric field does not affect the spin motion defined by (\ref{eq:g2}).  

Since 1975 all muon \g\ measurements have been made at the magic energy in continuous ring magnets with zero magnetic gradient and electric focusing.  The dilation of the muon lifetime to $64~\mu s$ has been an important factor in achieving the present accuracy, but we are now approaching the limits of this technique.  The $(g-2)$ frequency is determined over about 50 useful precession periods (three lifetimes) and departures from  a pure sine wave due to small phase and amplitude modulations \cite{benn02} can prejudice the fit.   Increasing the statistics to obtain better accuracy is not trivial and may augment the problem of signal overlap.  Furthermore, surveying the magnetic field to better than $0.1\ ppm$ with NMR probes carried around the ring is subject to uncertainties in calibration, alignment and field stablity.  

To improve the experiment it is desirable to increase the energy of the stored muons.  This would increase the lifetime so more \g\ cycles could be measured, giving a proportional increase in accuracy for the same number of stored particles.  But this would mean abandoning electric focusing, which can only be used at the magic energy of $3.1~GeV$.   It is also desirable to increase the magnetic field to speed up the precession and record more cycles per lifetime.  Two innovations make this possible: (a) a ring of discrete magnets with uniform field and edge focusing, separated by straight sections, the mean field over the orbit still being independent of radius; (b) calibration by protons in flight.

\begin{figure}[t!]
\includegraphics{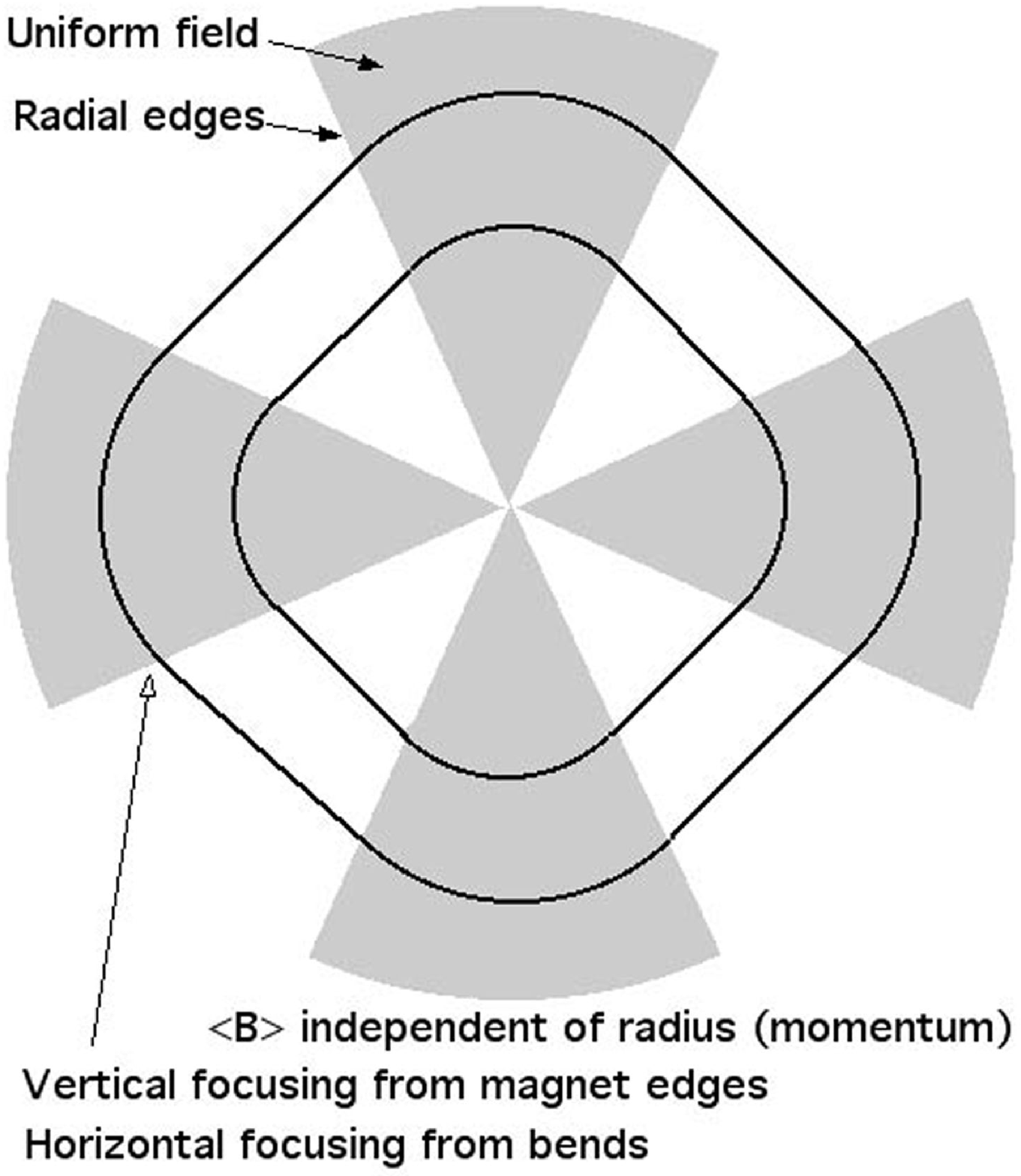}
\caption{\em Four-magnet ring.  In each magnet the particle is deflected through $90^o$.}
\label{fig1}
\end{figure}

The principle is illustrated in fig. \ref{fig1} which shows a ring with four bending magnets.  They are wedge shaped, with the magnet edges inclined along lines which are radial to the centre of the ring.  With four sectors this gives a quasi-square orbit with rounded corners.  If the track is scaled to a different size, the proportion of time spent inside the field is invariant; so the average magnetic field is independent of radius.  On the other hand, the particles cross the magnet edges at an angle, and this gives vertical focusing.  Uniform field magnets with edge focusing were used for the ZGS accelerator at Argonne \cite{symon56}.

Using standard formulae \cite{living} the horizontal and vertical tune, $Q_h$ and $Q_v$ can be calculated as a function of the edge angle.  With $N$ sectors the bend in each is $\psi = 2\pi /N$ .  Take the bending radius as 1 unit of length.  With straight sections of length $2\ x$ between sectors the edge angle $\theta $ is given by
\be
	T \equiv \tan {\theta} = x / \{ 1 + x \cot (\psi /2)\}	\label{eq;a}
\ee
implying
\be
	x = T/(1-Tz)   \label{eq:a1}
\ee
with  $z = cot(\psi /2)$.

In the horizontal plane, the matrix for the magnet edge is
$\left( \begin{array}{cc}
  1    &  0  \\
  T    &  1
\end{array} \right)$
and the matrix for the bend is    
$ \left( \begin{array}{rc}
 \: \: C     &   S \\
 -S     &  C
\end{array} \right)$
where $C = \cos {\psi }$ and $S = \sin {\psi }$.
The matrix for the straight section is  
$\left( \begin{array}{cc}
 1  &     2x  \\
 0   &    1
\end{array} \right)$
Multiplying out in the following order, edge, bend, edge, straight, we find that half the trace of the overall matrix for one sector is
\be
	H_h = C + ST + x\{ 2CT + S(T^2-1)\}			\label{eq:b}
\ee
The horizontal phase advance for one sector is given by the arc-cosine of half the trace of the matrix, so summing over all sectors the horizontal tune becomes
\be
	Q_h = \arccos (H_h) / \psi 					\label{eq:c}
\ee

\begin{figure}[t!]
\includegraphics{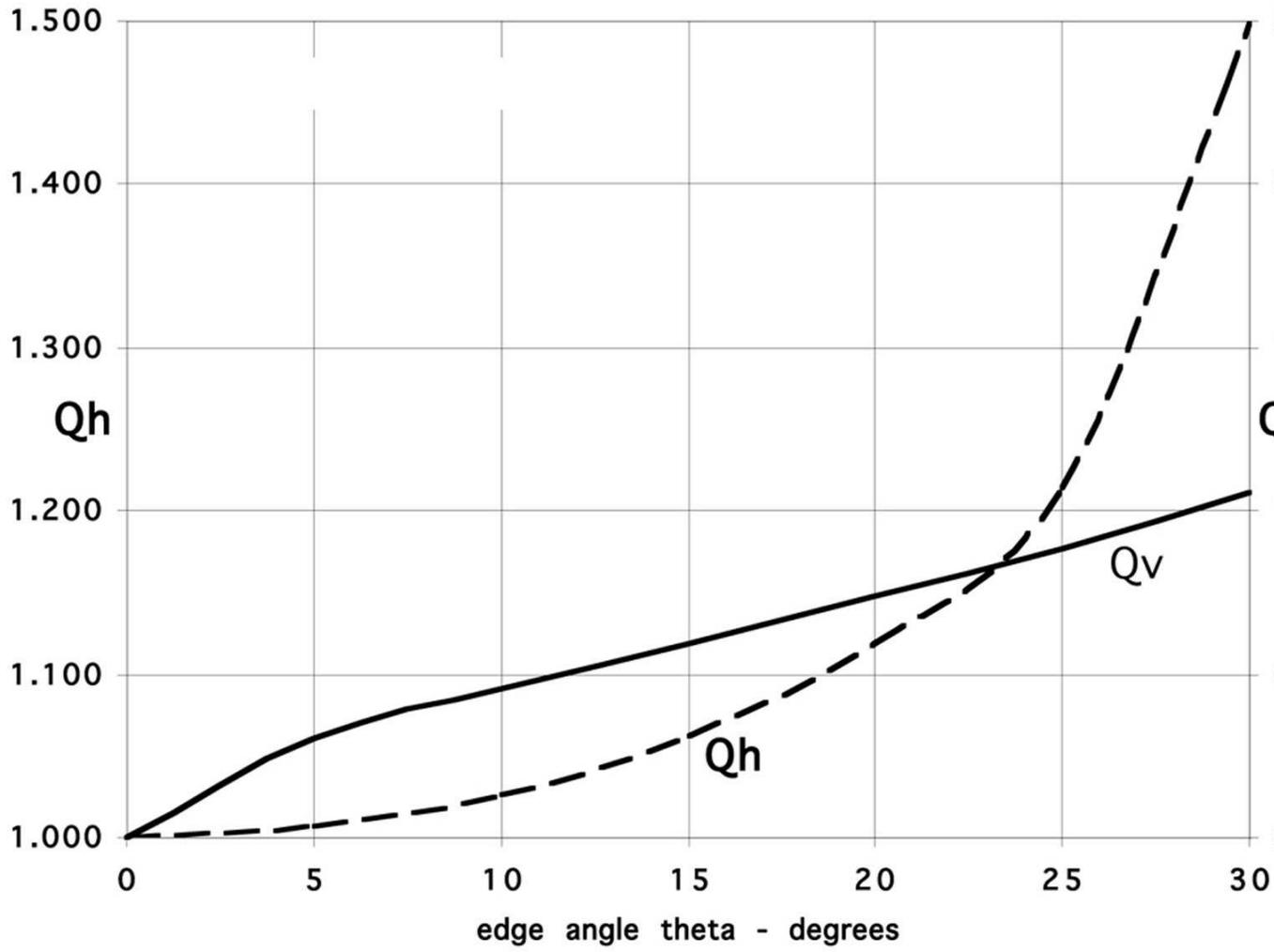}
\caption{\em Tune of three-magnet ring.  The bend in each magnet is $120^o$.  The edge angle is half the opening angle between magnets.}
\label{fig2}
\end{figure}

	In the vertical plane the calculation is similar, but the matrix for the edge is 
$\left( \begin{array}{rc}
  \: 1     &   0 \\
     -T     &  1
 \end{array} \right)$
while the matrix for the magnet is just a drift space of length $\psi $.   Multiplying out the matrices as before and using (\ref{eq:a1}), half the trace of the overall matrix for one sector in the vertical plane is
\be
	H_v = {1- Tz - 2T^2 - \psi T(1-Tz-2T^2)}/(1-Tz)	 \label{eq:d}
\ee
and the vertical tune is calculated from
\be
	Q_v = \arccos(H_v)/\psi                                   \label{eq:e} 
\ee
 With only two sectors, the orbit is unstable in the horizontal plane, but with three or more sectors many combinations of $Q_h$ and $Q_v$ can be obtained by varying $x$ or equivalently the bend angle  $\psi $ per sector.  With three sectors one obtains the highest average field for a given amount of focusing.  Fig. \ref{fig2} shows $Q_h$ and $Q_v$ as a function of edge angle in this case.

Calibration of the field cannot be achieved with standard NMR probes carried around the ring, because they will not function in the large gradients at the magnet edges.  Instead one can observe the precession of horizontally polarized protons in flight, which would automatically average the field over the orbit.  If the protons have the same momentum as the muons they must follow exactly the same track.  Orbit radii can be compared to sub-millimetre accuracy by measuring the rotation frequency of the particles while they are bunched \cite{brow00}.  (The small correction for the lower proton velocity can be applied with negligible error).

Horizontally polarized protons are regularly used \cite{ppol} in the Relativistic Heavy Ion Collider (RHIC) at Brookhaven and at $15\ GeV$ give an observed asymmetry of $\sim 0.6\% $ in scattering on a thin carbon ribbon.  The ribbon is $5\ \mu m$ wide and $3.5\ \mu g/cm^2$ ($150 \AA $) thick and the beam in the collider is observed for several periods of $20\ s$ with negligible loss of intensity, so deterioration of the beam by multiple scattering is not apparent.  The counting rate of the recoil carbon atoms is $\sim 1\ MHz$ for a stored beam of $\sim 2\times 10^{12}$ polarized protons.  In the proposed muon storage ring they would precess at the proton \g\ frequency, for example 100 $MHz$ in a field of $3.6\ T$.  As the average field is to be independent of radius, the spins would remain synchronized for at least $10^7$ turns of the spin ($\sim 100\ ms$).   As this ring would be $\sim 10$ times smaller than RHIC, by injecting for example $\sim 2\times 10^{11}$ particles for $100$ fills one would accumulate $10^7$ counts and the spin frequency would be measured in a few minutes to $\sim 7\ ppb$.  Probably one could use a thicker carbon ribbon without excessive scattering and get the same result with less injected beam.

The pitch correction \cite{pitch} to the \g\ frequency would desynchronize the spins if the pitch angle $\phi $ were too large.  But with $\phi $ expected to be $\sim 0.1\ mR$ the maximum correction ($\phi ^2/2$) is only $5\ ppb$.

The injected protons are bunched in time and because the momentum spread is very small they stay bunched for$\sim 1000$ turns.  The rotation frequency, seen in the detector, then measures the orbit radius to better than $1\ mm$.  The emittance of the proton beam is much smaller than that of the muons, so by varying the momentum it can be placed at various radii, and the field (averaged in azimuth) mapped as a function of radius.    In practice pole face windings would be used to make the final fine adjustments and ensure that the mean field was independent of radius.

In the vertical plane the protons oscillate about the median plane, sometimes missing the horizontal carbon ribbon, so this oscillation is seen in the detector.  By adjusting the injection conditions the oscillation can be minimized, so that the protons turn exactly in the median plane.   The beam is then located vertically by moving the carbon ribbon to maximize the signal.
 
 To map the field vertically one can change the median plane by introducing a radial magnetic field.  The best way to do this, without changing $B$, is to tilt the magnets!  For example, with $Q_v=0.4$ and mean radius $16~m$ tilting the magnets through an angle of $100~\mu R$  moves the median plane $10~mm$.  For measuring the tilt with nano-radian accuracy, the tiltmeter of R.V. Jones \cite{jones} is in use at Brookhaven \cite{bennup} and could be applied in a feedback loop to stabilize and control the tilt of the magnets.  In this way the proton beam can be used to map the field, averaged in azimuth, as a function of horizontal and vertical position in the aperture.  It will also be possible to inject the protons with a known vertical or horizontal betatron amplitude and so study how the average field for muons is influenced by oscillations about the equilibrium orbit.
 
Calibrated in this way, the experiment measures the ratio of muon to proton \g\ frequencies $f_a ^{\mu}/f_a ^p$.  To calculate the anomalous moment $a_\mu $ of the muon using Eq. (\ref{eq:g2calc}) one needs to convert the proton \g\ frequency $f_a ^p$ to the corresponding proton precession frequency at rest $f_s ^p$, that is one needs the ratio $Q=f_a ^p/f_s ^p = (g^p-2)/g^p$.  The gyromagnetic ratio $g^p$ of the proton is $2\times 2.792~ 847~ 386~ (63)$ \cite{pdg}, known to 23 $ppb$, so the error in the ratio $Q$ is only $8~ppb$ and the conversion from proton \g\ frequency to proton spin frequency at rest is sufficiently accurate.  By using protons in vacuum to calibrate the field one avoids all diamagnetic corrections due to molecular shielding \cite{diamag}, as well as questions of paramagnetic materials in or near the NMR probes.

A possible structure for 15 $GeV$ muons  comprises three 4.5 $T$ magnets with bend radius of 12 $m$ separated by straight sections 4.3 $m$ long.  The vertical tune would be $Q_v=0.4$, slightly higher than in the storage rings used at CERN and Brookhaven, and $Q_h=1.025$.  The average field would be $\approx 3.6~T$ and the accuracy for the same number of stored muons would be ten times better than at present.  The final parameters should be chosen to avoid orbit resonances and proton depolarization resonances.
One of the straight sections can be used for the proton polarimeter.  Another would have a conventional ferrite kicker for injecting the particles.  The residual field in the kicker should drop to low values after $\sim10~\mu s$ but in any case it could be tracked with the polarized protons.  As the incoming particles do not have to pass through a magnet, there would be no need for an inflector to cancel the magnet field along the incoming particle track \cite{krie89}.  

The advantages of this new ring structure are as follows:

  \begin{list}{}{\topsep-2mm \itemsep-1mm}
  \item[1)] No electric quadrupoles;
  \item[2)] No NMR trolley, which has to be calibrated, with corrections for the diamagnetism of water, etc;
  \item[3)] No superconducting inflector;
  \item[4)] Injection with a simple kicker in a field-free region; 
  \item[5)] Higher muon energy, giving longer lifetime, and proportionally higher accuracy.
  \item[6)]  Lower counting rates for the same number of stored muons, so less problem with overlapping signals;
  \item[7)] More time for detectors to recover from initial transients and particles created at injection;
  \item[8)] Higher magnetic field giving more \g\ cycles per lifetime and more accuracy: this is not possible with electric focusing as the electric field must be increased pro rata and is limited by breakdown.
 \end{list}
 
Muon sources of reasonable intensity are available from existing accelerators and an experiment at the level of $0.03~ppm$ (20 times better than at present) looks feasible.  Ensuring field stability while changing from proton to muon storage would require careful design.  Many details remain to be worked out; this is only a first report on a new concept.

This work was supported in part by the U.S. Department of Energy.  I am indebted to Professor Vernon Hughes at Yale for support and encouragement by over many years.  I thank G. Bunce and Paul T. Debevec for some helpful remarks.

\pagebreak

 \end{document}